\documentstyle[psfig]{mn} 
\title{The intermediate-age open clusters Ruprecht~61, 
Czernik~32, NGC~2225 and NGC~2262}
\author[G. Carraro et al.]        
{G. Carraro$^{1,2}$
\thanks{On leave from Dipartimento di Astronomia, Universit\`a di Padova,
Vicolo Osservatorio 2, I-35122, Padova, Italy.
email: gcarraro@das.uchile cl}, G. Baume$^{3}$, 
R.A.  V\'azquez$^{3}$, A. Moitinho$^{4}$, and D. Geisler$^{5}$ \\
$^1$Departamento de Astronom\'ia, Universidad de Chile, 
Casilla 36-D, Santiago, Chile\\
$^2$Astronomy Department, Yale University, 
P.O. Box 208101, New Haven, CT 06520-8101 , USA\\
$^3$Facultad de Ciencias Astron\'omicas y Geof\'{\i}sicas de la
UNLP, IALP-CONICET, Paseo del Bosque s/n, La Plata, Argentina\\
$^4$CAAUL, Observat\'orio Astron\'omico de Lisboa, Tapada da Ajuda,
  1349-018 Lisboa, Portugal\\
$^5$Universidad de Concepci\'on, Departamento de Fisica, 
Casilla 160-C, Concepci\'on, Chile\\          
} 
 
\date{\it Submitted: February 2005} 
\pubyear{2005} 
\begin{document} 
\maketitle 
\title{} 
 
%%%%%%%%%%%%%%%%%%%%%%%%%%%%%%%%%%%%%%%%%%%%%%%%%%%%%%%%% 
\begin{abstract} 
We present the first $BVI$ CCD photometry to $V=22.0$ of 4 fields centered  
on the region of the southern Galactic
star  clusters Ruprecht~61, Czernik~32, NGC~2225 and NGC~2262 
and of 4 displaced control fields. 
These clusters were never studied before, and we provide
for the first time estimates of their fundamental parameters,
namely radial extent, age, distance and reddening.
We find that the four clusters are all of intermediate age (around 1
Gyr), close to the Sun and possess lower than solar
metal abundance.
%\par 
\end{abstract} 
 

\section{Introduction}
This paper continues a series dedicated to the study of open clusters
in the third Galactic Quadrant, and aims at addressing
fundamental questions like the structure of the spiral arms in this quadrant,
and the precise  definition of the Galactic disk radial abundance gradient
outside the solar circle.
A more detailed illustration of the motivations of this project are given 
in Moitinho (2001) and Baume et al (2004).
Here we concentrate on four intermediate-age clusters
(about the age of the Hyades - 600 Myrs - or older)  
Ruprecht~61 (VdB-Hagen~32), Czernik~32 (VdB-Hagen~11), 
NGC~2225 and NGC~2262, for which
we provide the first photometric data and try to clarify
their nature and to derive estimates of their
fundamental parameters.\\
In Table~1 we list the cluster coordinates, which we redetermined 
from Digital Sky Survey (DSS) maps on a visual inspection basis.\\

\noindent
The layout of the paper is as follows. Sect.~2 illustrates  
the observation and reduction strategies. 
An analysis of  the geometrical
structure and star counts in the field of the clusters
is presented in Sect.~3, whereas a discussion of
the Color-Magnitude Diagrams (CMDs) is performed in Sect.~4.
Sect.~5 deals with the determination of clusters' reddening, 
distance, metallicity and age and,
finally, Sect.~6 summarizes our findings.

\begin{table}
\caption{Basic parameters of the clusters under investigation.
Coordinates are for J2000.0 equinox}
\begin{tabular}{ccccc}
\hline
\hline
\multicolumn{1}{c}{Name} &
\multicolumn{1}{c}{$RA$}  &
\multicolumn{1}{c}{$DEC$}  &
\multicolumn{1}{c}{$l$} &
\multicolumn{1}{c}{$b$} \\
\hline
& {\rm $hh:mm:ss$} & {\rm $^{o}$~:~$^{\prime}$~:~$^{\prime\prime}$} & [deg] & [deg]\\
\hline
Ruprecht~61          & 08:25:14 & -34:08:31 & 253.48 & +2.08\\ 
Czernik~32           & 07:50:30 & -29:50:36 & 245.86 & -1.74\\  
NGC~2225             & 06:26:37 & -09:38:07 & 218.78 & -9.86\\
NGC~2262             & 06:39:39 & +01:08:30 & 210.57 & -2.10\\
\hline\hline
\end{tabular}
\end{table}

\section{Observations and Data Reduction} 
 
$\hspace{0.5cm}$
CCD $BVI$ observations were carried out with the CCD camera on-board
the  1. 0m telescope at Cerro Tololo Interamerican Observatory (CTIO,Chile), on the nights of 
December 13 and 15, 2004. 
With a pixel size of $0^{\prime\prime}.469$,  and a CCD size of 512 $\times$ 512
pixels,  each pointing samples a $4^\prime.1\times4^\prime.1$ field on the sky.\\
\noindent
The details of the observations are listed in Table~2 where the observed 
fields are 
reported together with the exposure times, the average seeing values and the 
range of air-masses during the observations. 
Figs.~1 to 4 show I=600 secs images obtained in the area of 
Ruprecht~61, Czernik~32, NGC~2225 and NGC~2262, respectively.
Together with the clusters, we observed three control fields  15 arcmins
apart from the nominal cluster centers to deal with field star
contamination. Exposure of 600 secs in V and I were secured for these
fields.\\

\noindent
The data have been reduced with the 
IRAF\footnote{IRAF is distributed by NOAO, which are operated by AURA under 
cooperative agreement with the NSF.} 
packages CCDRED, DAOPHOT, ALLSTAR and PHOTCAL using the point spread function (PSF)
method (Stetson 1987). 
The two nights turned out to be photometric and very stable, and therefore
we derived calibration equations for all the 130 standard stars
observed during the two nights in the Landolt 
(1992)  fields SA~95-41, PG~0231+051, Rubin~149, Rubin~152,
T~phe and    SA~98-670 (see Table~2 for details).

 %%%%%%%%Figure 1  %%%% mappa 
\begin{figure} 
\centerline{\psfig{file=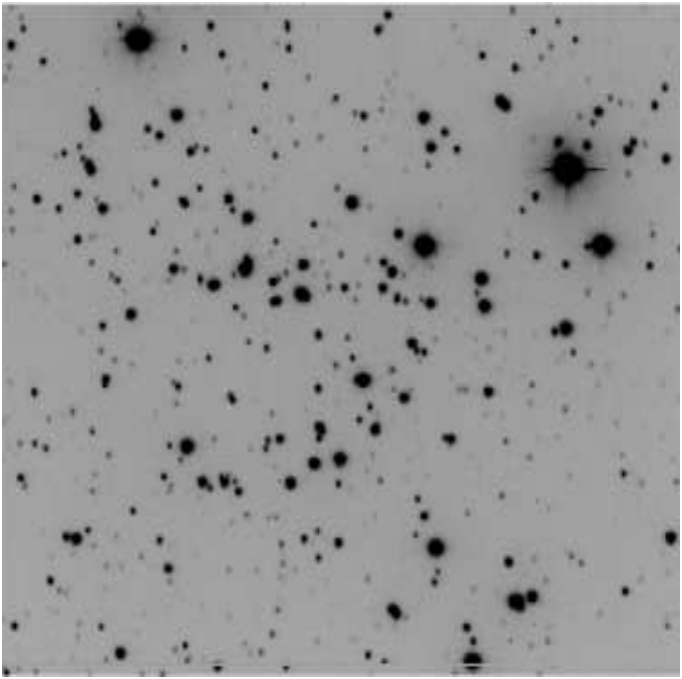,width=9.cm,height=9.cm}} 
\caption{Deep I image (600 secs)
of the open cluster Ruprecht~61. 
North is up, East on the left, and the covered area is $
4^{\prime}.1 \times 4^{\prime}.1$}
\label{mappa} 
\end{figure} 

\begin{figure} 
\centerline{\psfig{file=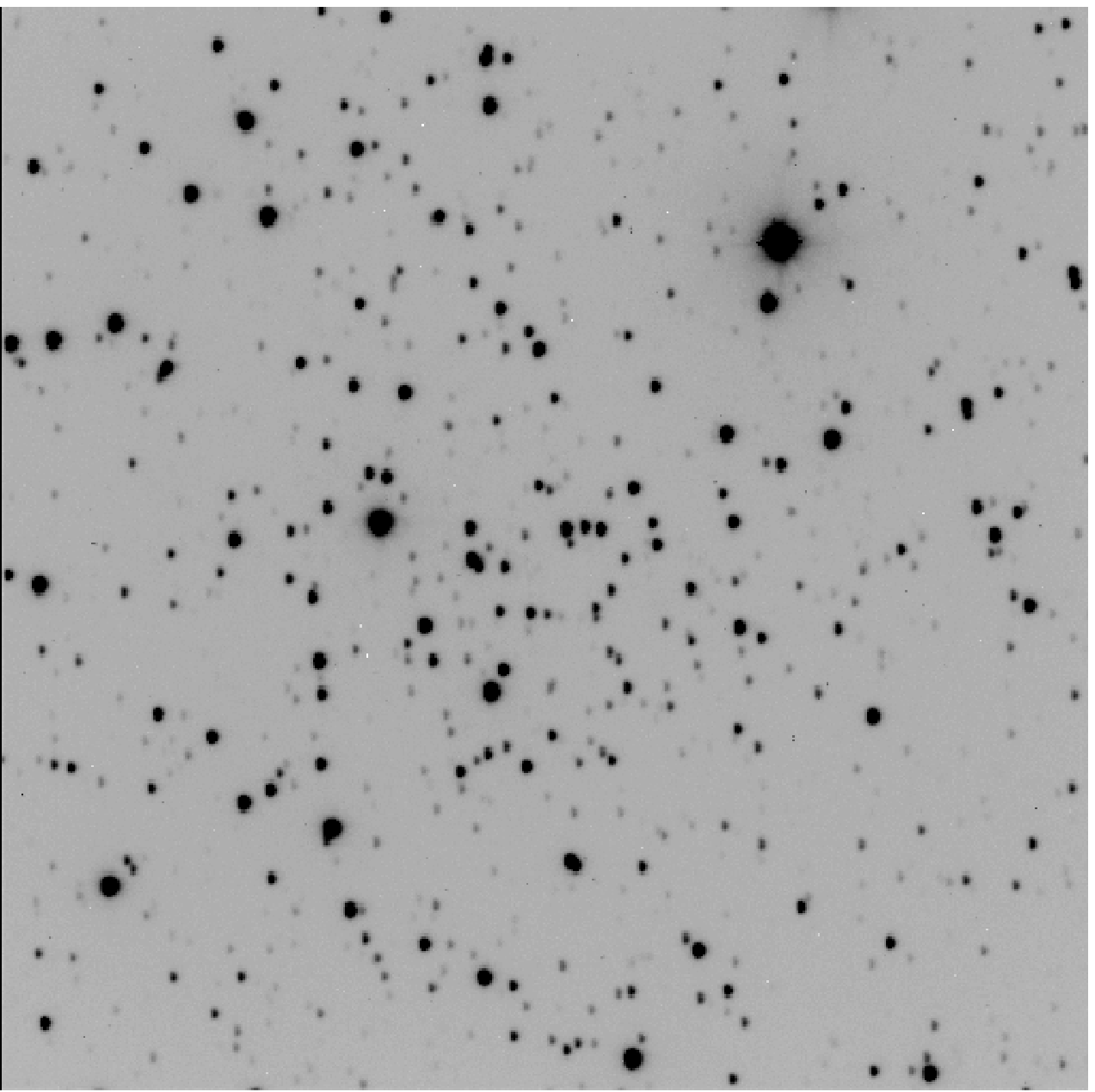,width=9.cm,height=9.cm}} 
\caption{Deep I image (600 secs)
of the open cluster Czernik~32. 
North is up, East on the left, and the covered area is $
4^{\prime}.0 \times 4^{\prime}.1$}
\label{mappa} 
\end{figure}

\begin{figure} 
\centerline{\psfig{file=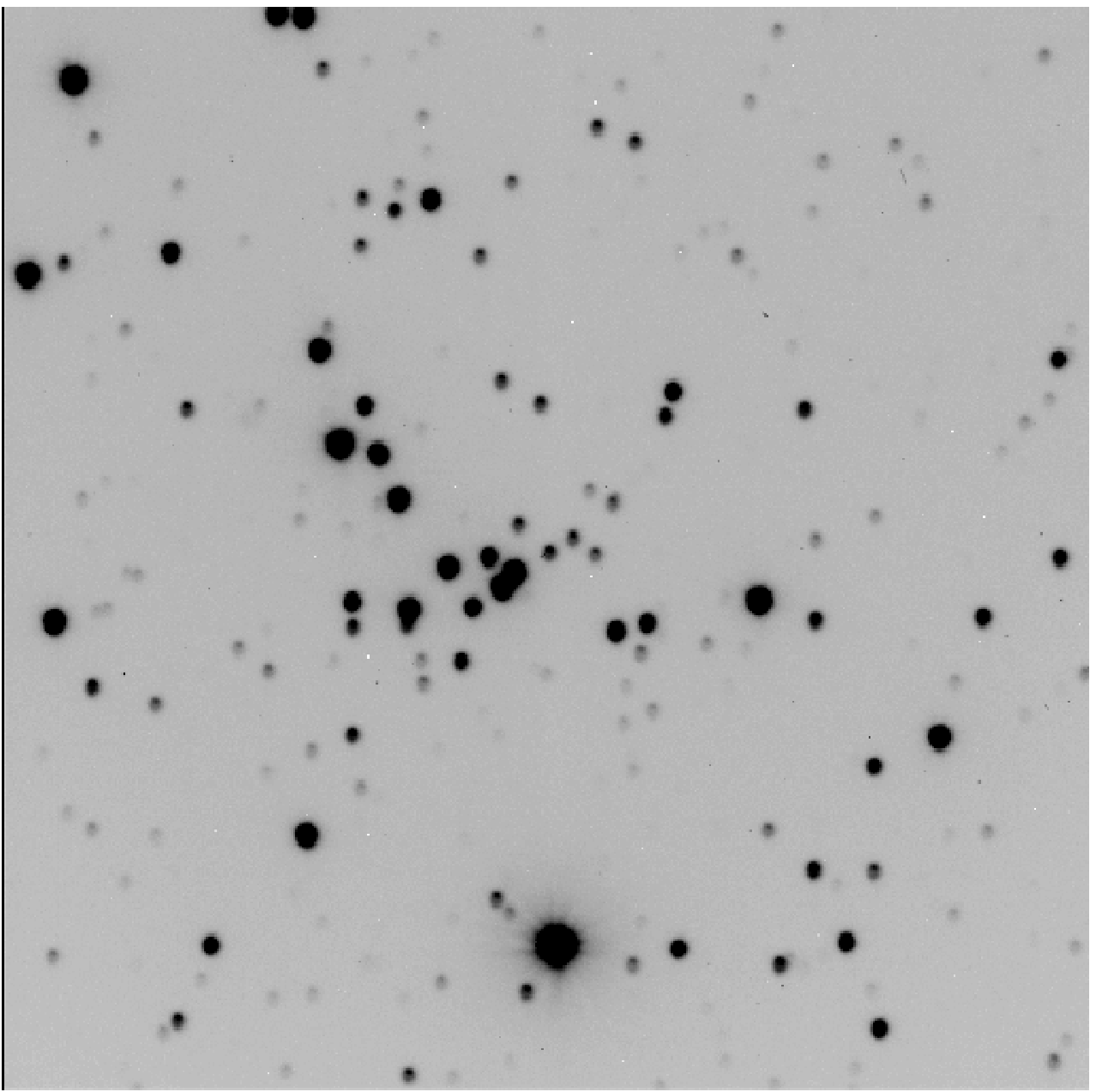,width=9.cm,height=9.cm}} 
\caption{Deep I image (600 secs)
of the open cluster NGC~2225. 
North is up, East on the left, and the covered area is $
4^{\prime}.1 \times 4^{\prime}.1$}
\label{mappa} 
\end{figure}

\begin{figure} 
\centerline{\psfig{file=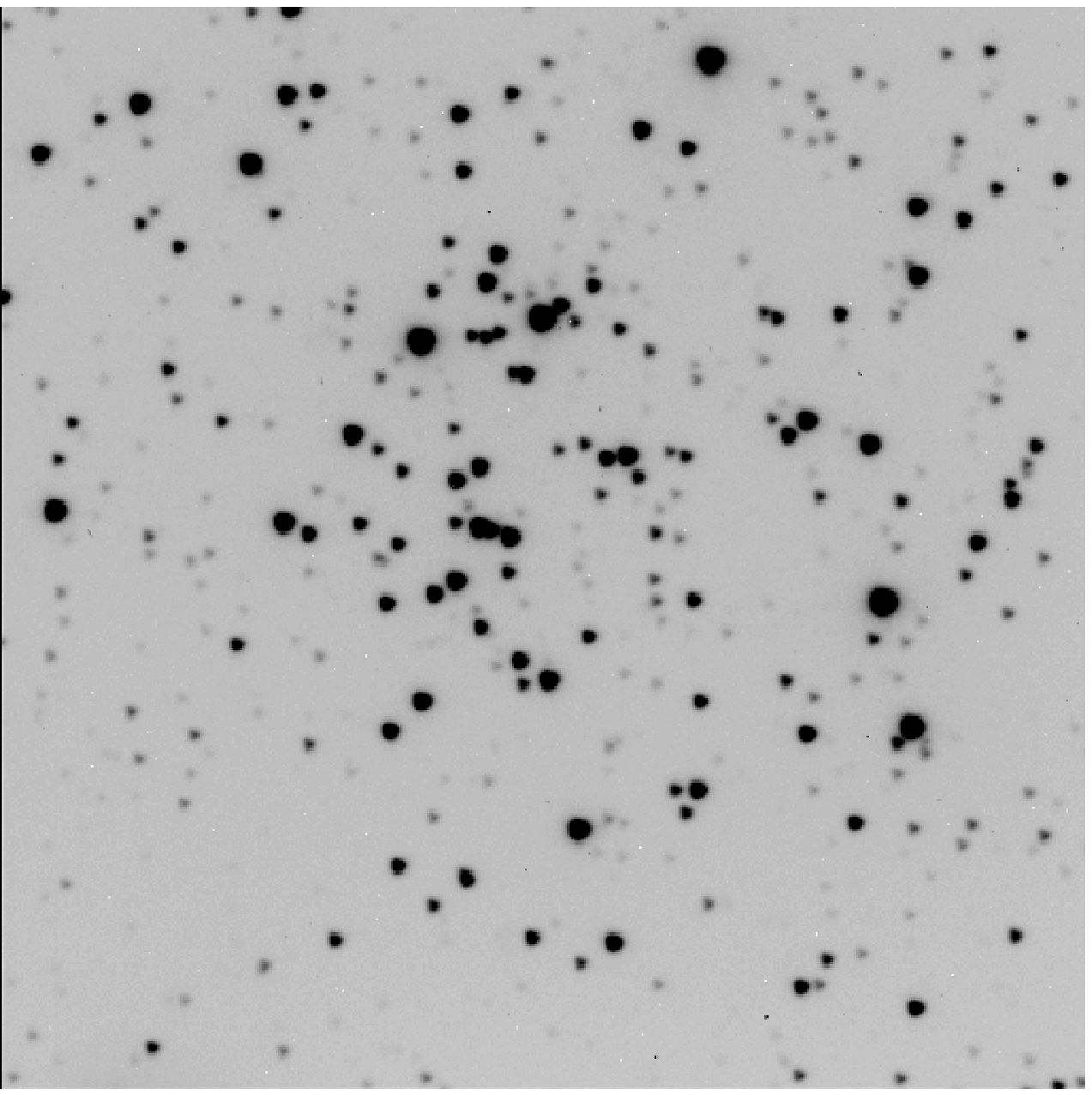,width=9.cm,height=9.cm}} 
\caption{Deep I image (600 secs)
of the open cluster NGC~2262. 
North is up, East on the left, and the covered area is $
4^{\prime}.1 \times 4^{\prime}.1$}
\label{mappa} 
\end{figure}

\begin{table} 
\fontsize{8} {10pt}\selectfont
\tabcolsep 0.10truecm 
\caption{Journal of observations of Ruprecht~61, Pismis~7 and Czernik~32 
and standard star fields (December 13 and 15, 2004).} 
\begin{tabular}{cccccc} 
\hline 
\multicolumn{1}{c}{Field}         & 
\multicolumn{1}{c}{Filter}        & 
\multicolumn{1}{c}{Exposure time} & 
\multicolumn{1}{c}{Seeing}        &
\multicolumn{1}{c}{Airmass}       \\
 & & [sec.] & [$\prime\prime$] & \\ 
\hline 
Ruprecht~61    & B &         120,1200   &   1.2 & 1.00-1.10 \\
              & V &      30,600    &   1.3 & 1.00-1.10 \\ 
              & I &      30,600    &   1.2 & 1.00-1.10 \\
\hline
NGC~2262     & B &         120,1200   &   1.2 & 1.15-1.30 \\
              & V &      30,600    &   1.3 & 1.15-1.30 \\ 
              & I &      30,600    &   1.2 & 1.15-1.30 \\
\hline
NGC~2225     & B &         120,1200   &   1.2 & 1.15-1.30 \\
              & V &      30,600    &   1.3 & 1.15-1.30 \\ 
              & I &      30,600    &   1.2 & 1.15-1.30 \\
\hline
Czernik~32     & B &         120,1200   &   1.2 & 1.25-1.40 \\
              & V &      30,600    &   1.3 & 1.25-1.40 \\ 
              & I &      30,600    &   1.2 & 1.25-1.40 \\
\hline
SA 98-670     & B &   $3 \times$120   &   1.2 & 1.24-1.26 \\
              & V &   $3 \times$40    &   1.4 & 1.24-1.26 \\ 
              & I &   $3 \times$20    &   1.4 & 1.24-1.26 \\ 
\hline
PG 0231+051   & B &   $3 \times$120   &   1.2 & 1.20-2.04 \\
              & V &   $3 \times$40    &   1.5 & 1.20-2.04 \\ 
              & I &   $3 \times$20    &   1.5 & 1.20-2.04 \\ 
\hline
T Phe         & B &   $3 \times$120   &   1.2 & 1.04-1.34 \\
              & V &   $3 \times$ 40   &   1.3 & 1.04-1.34 \\ 
              & I &   $3 \times$ 20   &   1.3 & 1.04-1.34 \\ 
\hline
Rubin 152     & B &   $3 \times$120   &   1.3 & 1.33-1.80 \\
              & V &   $3 \times$40    &   1.2 & 1.33-1.80 \\ 
              & I &   $3 \times$20    &   1.2 & 1.33-1.80 \\ 
\hline
Rubin 149     & B &   $3 \times$120   &   1.1 & 1.21-1.96 \\
              & V &   $3 \times$40    &   1.2 & 1.21-1.96 \\ 
              & I &   $3 \times$20    &   1.2 & 1.21-1.96 \\ 
\hline
SA 95-41      & B &   $3 \times$120   &   1.2 & 1.05-1.48 \\
              & V &   $3 \times$40    &   1.2 & 1.05-1.48 \\ 
              & I &   $3 \times$20    &   1.1 & 1.05-1.48 \\ 
\hline
\hline
\end{tabular}
\end{table}

\noindent
The adopted calibration equations are of the form:\\

\noindent
$ b = B + b_1 + b_2 \times X + b_3 \times(B-V)$ \\
$ v = V + v_1 + v_2 \times X + v_3 \times(B-V)$ \\
$ v = V + v_{1,i} + v_{2,i} \times X + v_{3,i} \times (V-I)$ \\
$i = I + i_1 + i_2 \times  X + i_3 \times (V-I)$ ,\\

\begin{table} 
\tabcolsep 0.3truecm
\caption {Coefficients of the calibration equations}
\begin{tabular}{ccc}
\hline
$b_1 = 3.465 \pm 0.009$ & $b_2 =  0.25 \pm 0.02$ & $b_3 = -0.145 \pm 0.008$ \\
$v_1 = 3.244 \pm 0.005$ & $v_2 =  0.16 \pm 0.02$ & $v_3 =  0.021 \pm 0.005$ \\
$v_{1,i} = 3.244 \pm 0.005$ & $v_{2,i} =  0.16 \pm 0.02$ & $v_{3,i} =  0.009 \pm 0.005$ \\
$i_1 = 4.097 \pm 0.005$ & $i_2 =  0.08 \pm 0.02$ & $i_3 =  0.006 \pm 0.005$ \\
\hline
\end{tabular}
\end{table}

\noindent
where $BVI$ are standard magnitudes, $bvi$ are the instrumental ones and  $X$ is 
the airmass; all the coefficient values are reported in Table~3.
The standard 
stars in these fields provide the color coverage
$-0.6 \leq (B-V) \leq 2.2)$.
The final {\it r.m.s.} of the calibration are 0.031, 0.024 and 0.023 for the B, V and I filters,
respectively.
\noindent
We generally used the third equation to calibrate the $V$ magnitude
in order to get the same magnitude depth both in the cluster
and in the field.
\noindent
Photometric errors have been estimated following closely Patat \& Carraro (2001, Appendix A),
which the reader is referred to for all the details.
It turns out that the   
global photometric errors amount to
0.03, 0.05 and 0.20 at V = 12, 16 and 21.5 mag, respectively.\\
\noindent
The final photometric catalogs for Ruprecht~61, Czernik~32, NGC~2225 and NGC~2262 
(coordinates, B, V and I magnitudes and errors)  
consist of 1166, 1047, 869 and 925 stars, respectively, and are made 
available in electronic form at the  
WEBDA\footnote{http://obswww.unige.ch/webda/navigation.html} site
maintained by J.-C. Mermilliod.\\

\begin{figure} 
\centerline{\psfig{file=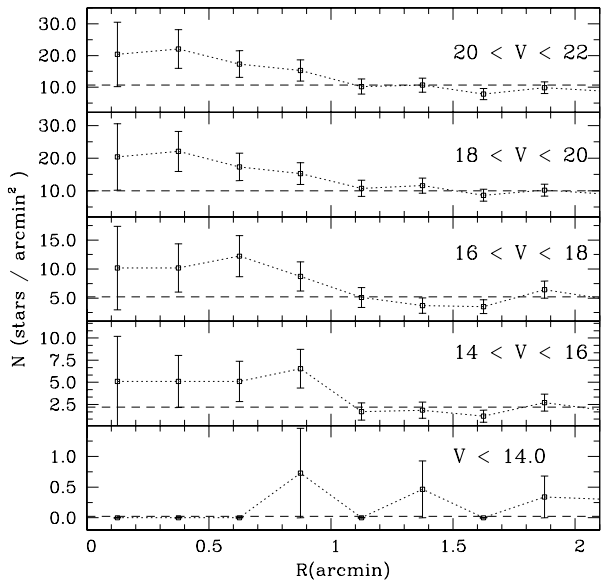,width=\columnwidth}} 
\caption{Star counts in the area of 
Ruprecht~61 as a function of  radius and magnitude. The dashed lines represent
the level of the control field counts estimated from the accompanying control field.}
\end{figure}

\section{Star counts and clusters' sizes} 
As we will show in this Section, 
our photometry covers entirely each cluster's area
allowing us to  perform star counts to obtain
improved estimates of the clusters' size.
In fact these clusters are generally very faint, poorly populated  and compact 
(see Fig~.1 to 4) and therefore could well fit within the CCD area.\\

\noindent
We derived the surface stellar density by performing star counts
in concentric rings around the clusters' nominal centers (see Table~1)
and then dividing by their
respective areas. Poisson errors have also been derived and normalized
to the corresponding area. 
The field star contribution has been derived from the control
field which we secured for each cluster, and the errors have been
computed in the same way as for the cluster field.\\

\begin{figure}
\centerline{\psfig{file=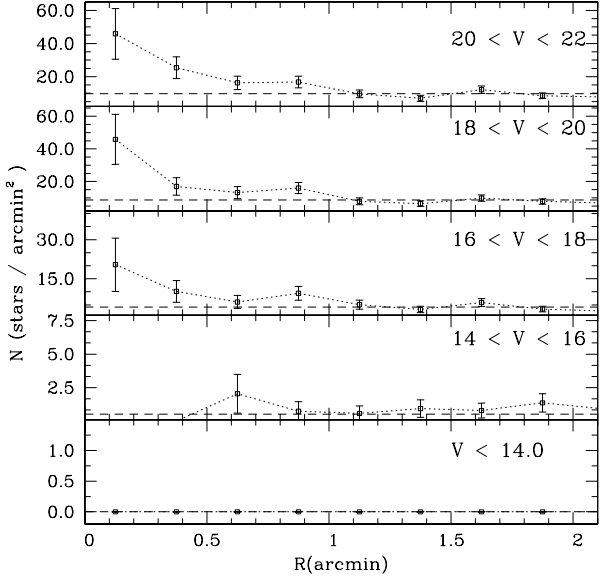,width=\columnwidth}} 
\caption{Star counts in the area of 
Czernik~32 as a function of  radius and magnitude. The dashed lines represent
the level of the control field counts estimated from the
accompanying control field.}
\end{figure}

\begin{figure}
\centerline{\psfig{file=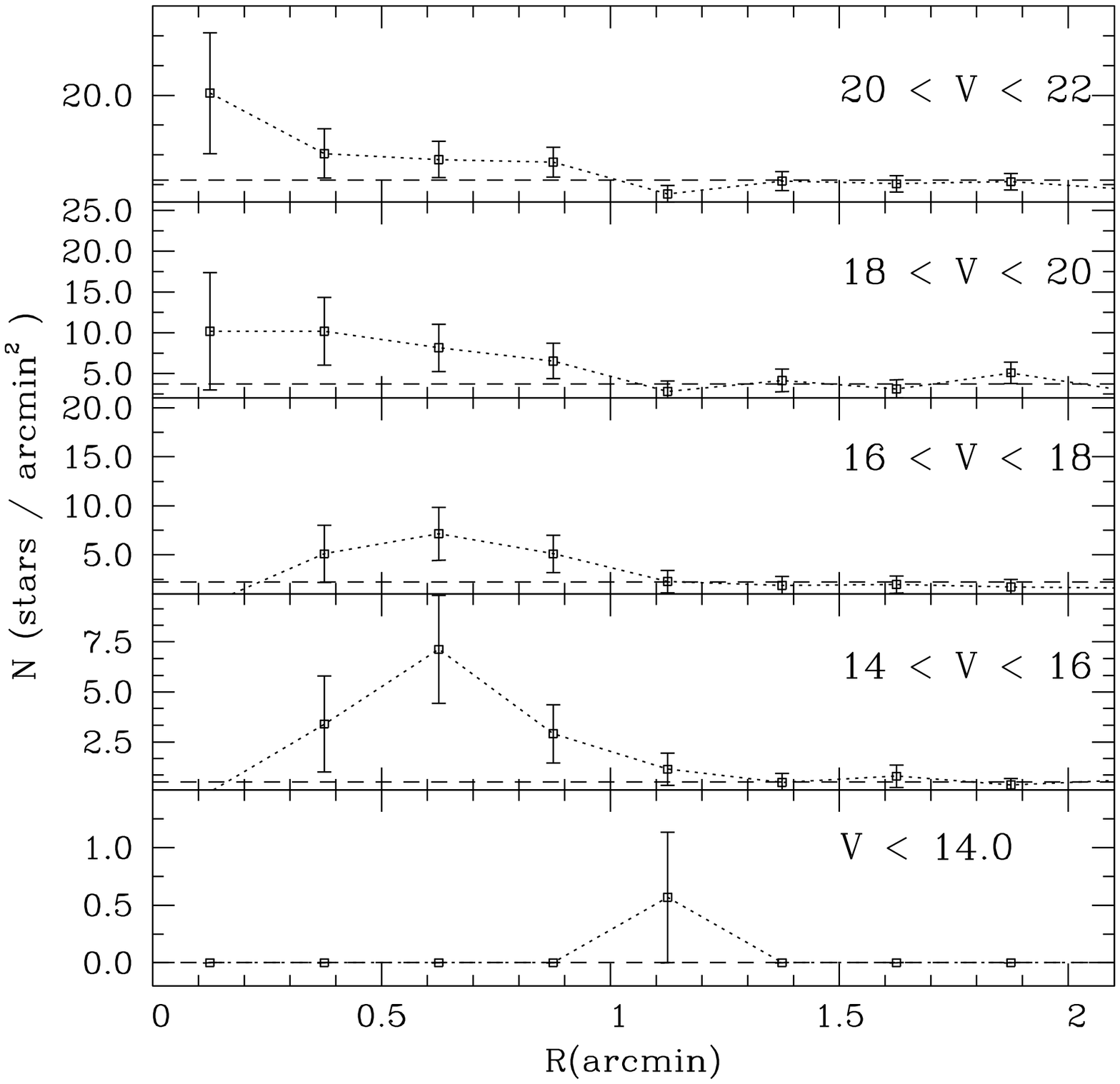,width=\columnwidth}} 
\caption{Star counts in the area of 
NGC~2225 as a function of  radius and magnitude. The dashed lines represent
the level of the control field counts estimated from the
accompanying control field.}
\end{figure}

\begin{figure}
\centerline{\psfig{file=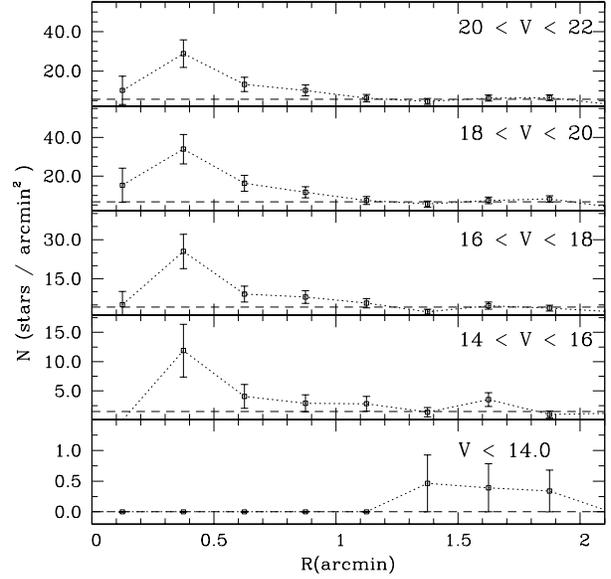,width=\columnwidth}} 
\caption{Star counts in the area of 
NGC~2262 as a function of  radius and magnitude. The dashed lines represent
the level of the control field counts estimated from the
accompanying control field.}
\end{figure}

\noindent
{\bf Ruprecht~61}
The radial density profile for Ruprecht~61 is shown in Fig.~5
as a function of the V magnitude.
Clearly,  the cluster does not appear very concentrated, and it is
deficient of bright stars.
The cluster seems to emerge from the background
in the magnitude range
$16 \leq V \leq 22$.
In this magnitude range the radius is not larger than 1.0 arcmin.
\noindent
We shall adopt the  value of 1.0 arcmin as the radius 
of Ruprecht~61
throughout this paper. This estimate is in perfect agreement
the value of 2.0 arcmin reported by Dias et al. (2002) for the cluster
diameter.\\

\noindent
{\bf Czernik~32}
The radial density profile for Czernik~32 is shown in Fig.~6
as a function of the V magnitude.
The cluster shows a clear lack of bright stars ant it is mostly
populated  by stars of magnitude in the range
$18 \leq V \leq 22$.
In this magnitude range the radius is not larger than 1.0 arcmin.
\noindent
In conclusion,  we are going to adopt the  value of 1.0 arcmin as 
the Czernik~32
radius throughout this paper. This estimate is in reasonable agreement with
the value of 3.0 arcmin reported by Dias et al. (2002) for the cluster
diameter.\\

\noindent
{\bf NGC~2225}
The radial density profile for NGC~2225 is shown in Fig.~7
as a function of the V magnitude.
Also this cluster exhibits a deficiency of bright stars ant it is mostly
populated  by stars of magnitude in the range
$16 \leq V \leq 22$.
In this magnitude range the radius is not larger than 1.2 arcmin.\\
\noindent
This estimate is in reasonable agreement with (a bit smaller than) 
the value of 4.0 arcmin reported by Dias et al. (2002) for the cluster
diameter.\\

\noindent
{\bf NGC~2262}
The  radial density profile for NGC~2262 is shown in Fig.~8
as a function of the V magnitude.
This cluster is composed mainly
by stars of magnitude in the range
$16 \leq V \leq 22$, where
the radius is around 1.0 arcmin.
\noindent
In conclusion,  we are going to adopt the  value of 1.0 arcmin as 
the radius of  NGC~2262
throughout this paper. This estimate is much smaller than
the value of 5.0 arcmin reported by Dias et al. (2002) for the cluster
diameter.\\

\noindent
The estimates we provide for the radius, although reasonable,
must be taken as preliminary.
In fact, the size of the CCD is probably too small to derive 
conclusive estimates of the cluster sizes, that due to dynamical evolution
and mass segregation tend to be normally
under-estimated. 
Larger and deep field coverage
is necessary to derive firmer estimates of the clusters
radii.

\begin{figure*} 
\centerline{\psfig{file=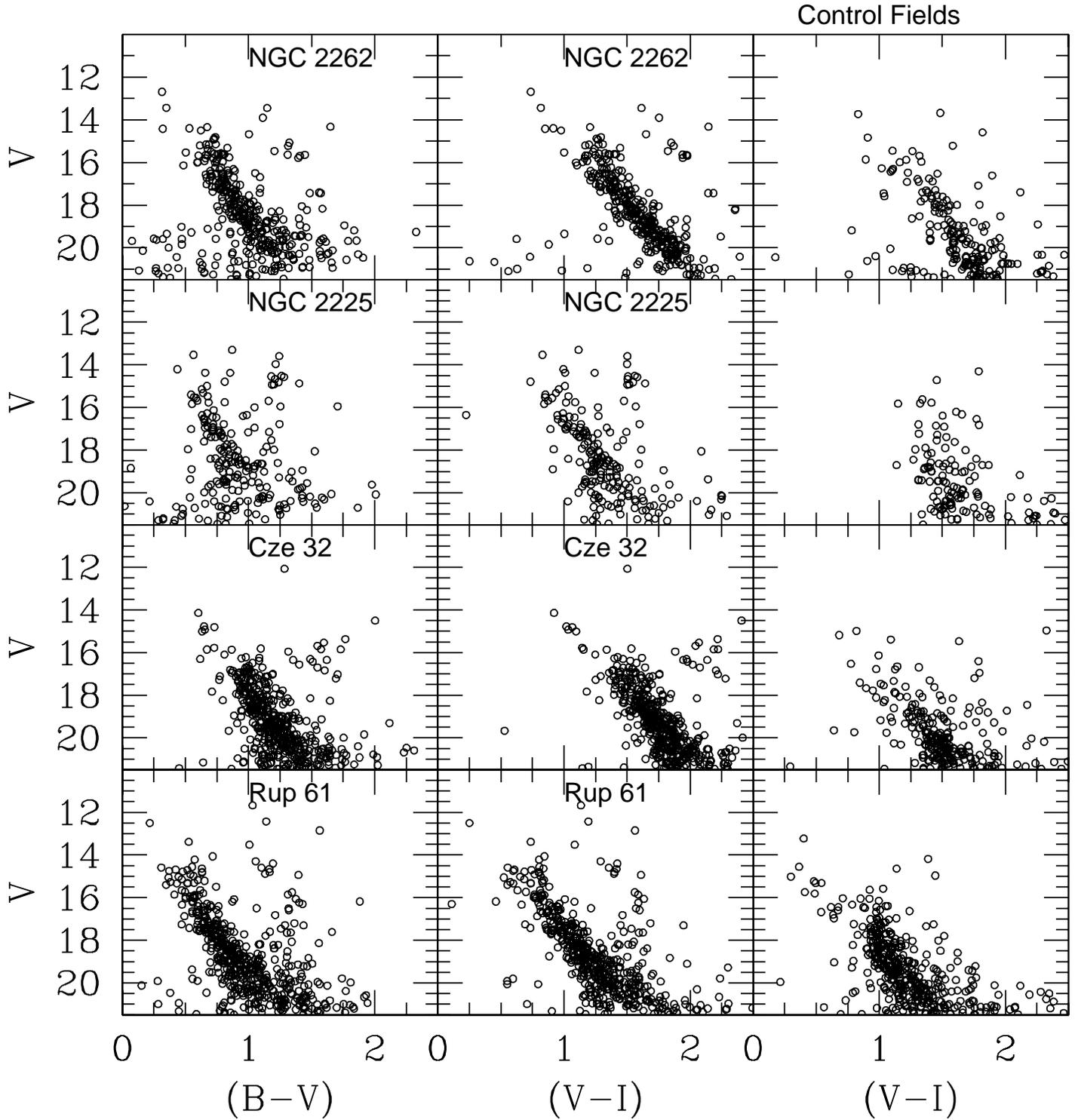}} 
\caption{$V$ vs $(B-V)$ (left panels) and $V$ vs $(V-I)$ (middle panels)
CMDs
of Ruprecht~61, Czernik~32, NGC~2225 and NGC~2262 
and corresponding control fields (right panels).
We include all stars in each field.}
\end{figure*} 

%%%%%%%%%%%%%%%%%%%%%%%%%%%%%%%%%%%%%%%%%%%%%%%%%%%%%%%%%%%%%%%%%%%%%%%%%%%%%%%%%%

\section{The Colour-Magnitude Diagrams} 
In Fig.~9 we present the CMDs obtained for 
the observed fields of the four clusters
under investigation.
All the stars observed in each field have been
plotted (not only those within the derived cluster
radii).
In this figure, the open cluster Ruprecht~61 is shown  together
with the corresponding control field in the lower panels,
Czernik~32 and NGC~2225 are presented in the
middle panels, while finally NGC~2262 is presented in the upper panels. 
The control fields
help us to better interpret these CMDs, which are clearly
affected by strong foreground star contamination.\\

\noindent
{\bf Ruprecht~61}.
This cluster is presented in the lower panels of Fig.~9.
It exhibits a Main Sequence (MS) extending from
V=15-15.5, where the Turn Off Point (TO) is located, down
to V=21.5. This MS is significantly wide, wider
than photometric error at a given magnitude (see Sect.~2).
We ascribe this to
field star contamination, 
and to the presence of a sizeable binary star population,
which mainly enlarge the MS toward red colors.
\noindent
However, the reality of this cluster seems to be secured by the shape of
the MS with respect to the control field MS, whose
population sharply decreases at V = 17.
Also, the cluster MS is significantly bluer and more tilted
than the field MS, which derives from the superposition
of stars of different reddening located at all distances between
the cluster and the Sun.
Another interesting evidence is the possible presence of  a clump
of stars at V=14.5, which does not have a clear counterpart
in the field, and which implies a cluster of intermediate-age.
In fact if we use the age calibration from Carraro \& Chiosi (2004),
for a $\Delta V$ (the magnitude difference between
the red clump and the TO) of 0.5 mag, we infer an age around 1 
Gyr. This estimate does not take into account the cluster metallicity,
and therefore is simply a guess. In the following Sect. we shall
provide a more robust estimate of the age through a detailed
comparison with theoretical isochrones.\\

\noindent
{\bf Czernik~32}.
The open cluster Czernik~32 is presented in the lower-mid panels
of Fig.~9.
The TO located at $V \approx$ 17, and a
prominent clump at $V \approx$ 16
with no counterpart in the field CMD are readily seen, yielding an estimated 
age of around 1.0 Gyr.
The overall morphology of the CMDs is in this case
very different from the
field CMD 
leaving no doubt that Czernik~32
is a bona-fide intermediate-age
open cluster.\\

\noindent
{\bf NGC~2225}.
The open cluster NGC 2225 is presented in the upper-mid panels
of Fig. 9.  Again, the overall morphology of the cluster CMDs, with a
TO and an evident red clump, is very different from the field CMD.
Indeed, the field sequence is much less populated and stops at V
$\approx$ 16.5, giving a first impression that this is a bona-fide
intermediate-age open cluster. The TO located at V $\approx$ 16, and
red clump at V $\approx$ 15, allow to estimate an age of around 1.0
billion years, confirming the first impression that NGC 2225 is an
intermediate age cluster.

\noindent
{\bf NGC~2262}.
The open cluster NGC~2262 is presented in the upper panels
of Fig.~9.
The cluster's CMD reveals a TO is located at $V \approx$ 16, 
and a possible  clump
at $V \approx$ 16 as well, which provide a rough estimate 
of 
around 0.5 Gyrs for the age of NGC~2262.
The overall morphology of the CMDs is also in this case 
very different from the
field CMD 
confirming that this is a bona-fide intermediate-age
open cluster.

\begin{figure*} 
\centerline{\psfig{file=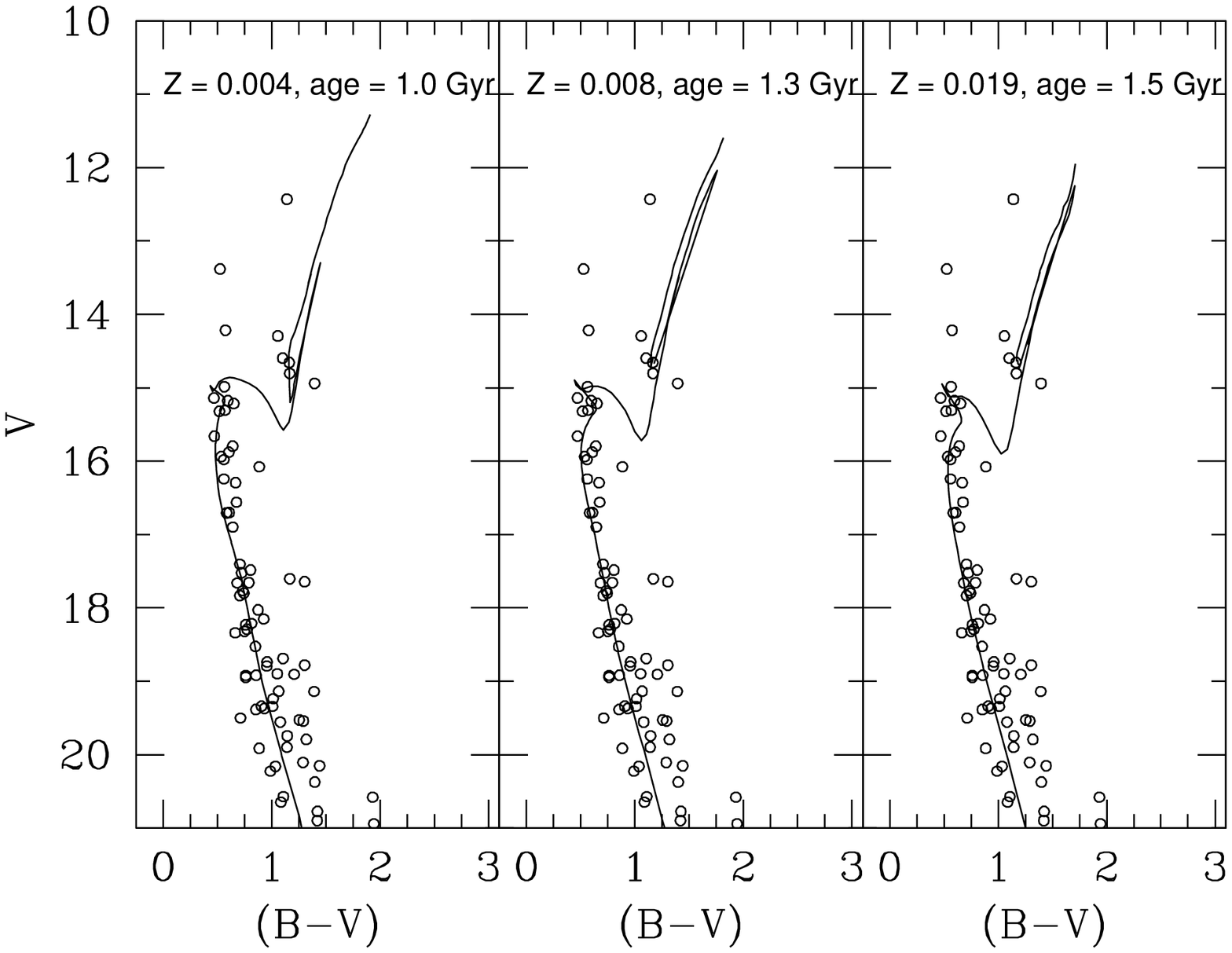}} 
\caption{Hunting for the best fit isochrone solution.
The CMD of Ruprecht~61 is shown together with
three different isochrones.
The isochrones
are for the age of 1 Gyr and metallicity Z=0.004
in the left panel, for the age of 1.3 Gyrs and metallicity Z=0.008
in the middle panel, and, finally, for the age of 1.5 Gyrs and metallicity Z=0.019
in the right panel.}
\end{figure*} 

\section{Deriving clusters' fundamental parameters}
In this section we are going to perform a detailed comparison
of the star distribution in the clusters' CMDs with theoretical
isochrones. For this study, we adopt in this study the Padova library from
Girardi et al. (2000).
This comparison is clearly not an easy exercise. In fact,
the detailed shape and position of the various features
in the CMD (MS, TO and clump basically) depends mostly
on age and metallicity, and then also on reddening and distance.
The complex interplay between the various parameters is however well
known, and we refer to 
Chiosi et al. (1992) and Carraro (2005)
as nice examples of the underlying technique.\\
Our basic strategy is to survey different age and metallicity
isochrones in an attempt to provide the best fit of all the CMD
features both in the $V$ vs $(B-V)$ and in the $V$ vs $(V-I)$ CMD.\\
To further facilitate the fitting procedure,
by increasing the contrast between the cluster and the field
population,
we shall consider only the stars which lie  within
the cluster radius as derived in Sect.~3.\\

\noindent
Finally, to derive the clusters' distances from reddening and apparent distance modulus,
a reddening law must be specified. In this study we shall adopt the 
normal reddening law
$Av = 3.1 \times E(B-V)$ in deriving the clusters' distances. \\

\noindent
Additionally to finding the best fit, we also estimated the  
uncertainties in the basic parameters. These uncertainties simply
reflect the range of parameters 
that yields a reasonable fit to the clusters CMDs.
The errors are reported in Table~4, and an example of the procedure is shown in Fig.10 for the case of Ruprecht~61.
The best fir for all the clusters,
achieved simultaneously in the V vs $(B-V)$ and in the $V$ vs $(V-I)$ planes,
are shown in Figs.~11-14.\\

\noindent
{\bf Ruprecht~61}.
The fitting procedure and the isochrone solution for this cluster are
shown in Figs.~10 and 11.
We obtained the best fit for an age of 1.3 Gyrs and a metallicity
Z=0.008 (see middle panel of Fig.~11, and Fig.~12).
In fact, the shape of the TO in the left panel of
Fig.~11 (for the Z=0.004 isochrone) is clearly different from the underlying cluster sequence,
and the same can be noticed for the Z=0.019 isochrone (right panel), where the 
red hook shows a shape which does 
not fit very well the star distribution
in the cluster. In details, the red hook is too red and somewhat
faint with respect to the Z=0.008 isochrone and the actual stars distribution. 
To get a bluer and brighter red hook, one should use a 
younger isochrone, which will however possess a too red clump and RGB, when
fixed to the TO.\\
The inferred
reddening and apparent distance modulus are E(B-V)=0.30 (E(V-I)=0.41, right panel in Fig.~12)
and (m-M)=13.85, respectively. As a consequence, the cluster
possesses a heliocentric distance of 3.9 kpc, and
is located at a Galactocentric distance of 9.4 kpc, assuming
8.5 kpc as the distance of the Sun to the Galactic Center.\\

\noindent
{\bf Czernik~32}.
The isochrone solution for this cluster is displayed in Fig.12.
We obtained the best fit for an age of 1 Gyr and a metallicity
Z=0.008. The inferred
reddening and apparent distance modulus are E(B-V)=0.85 (E(V-I)=1.08)
and (m-M)=15.7, respectively. These values situate the cluster at a heliocentric distance
of 4.1 kpc,
which corresponds
to a Galactocentric distance of 10.8 kpc. 
The overall fit is also good in this case, the detailed shape of the MS
and TO are nicely reproduced, as well as the color of the clump.\\

\noindent
{\bf NGC~2225}.
The isochrone solution for this cluster is presented in Fig.13.
We obtained the best fit for an age of 1 Gyr and a metallicity
Z=0.008, which reproduces the sharp cluster sequence
extremely well. The inferred
reddening and apparent distance modulus are E(B-V)=0.35 (E(V-I)=0.50)
and (m-M)=13.6, respectively. Therefore the cluster has a heliocentric distance
of 3.2 kpc,
and
is located at a Galactocentric distance of 11.2 kpc. \\

\noindent
{\bf NGC~2262}.
The isochrone solution for this cluster is shown in Fig.14.
We obtained the best fit for an age of 1 Gyr and a metallicity
of Z=0.008. The inferred
reddening and apparent distance modulus are E(B-V)=0.55 (E(V-I)=0.72)
and (m-M)=14.5, respectively, which put the cluster at a heliocentric distance
of 3.6 kpc,
or 
at a Galactocentric distance of 11.7 kpc.

\begin{figure} 
\centerline{\psfig{file=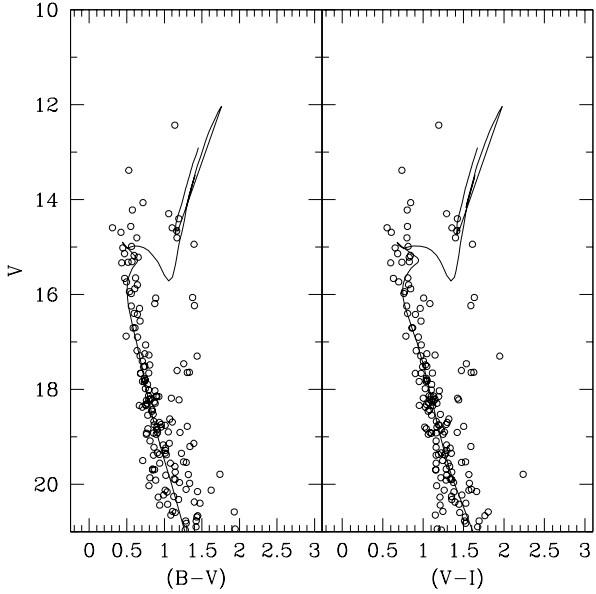,width=\columnwidth}} 
\caption{Isochrone solution for Ruprecht~61. The isochrone
is for the age of 1.3 Gyr and metallicity Z=0.008.
The apparent distance modulus is (m-M)=13.85, and the reddening
E(B-V)=0.30 and E(V-I)=0.41. See text for more details. Only stars within the derived radius are shown. }
\end{figure}

\begin{figure} 
\centerline{\psfig{file=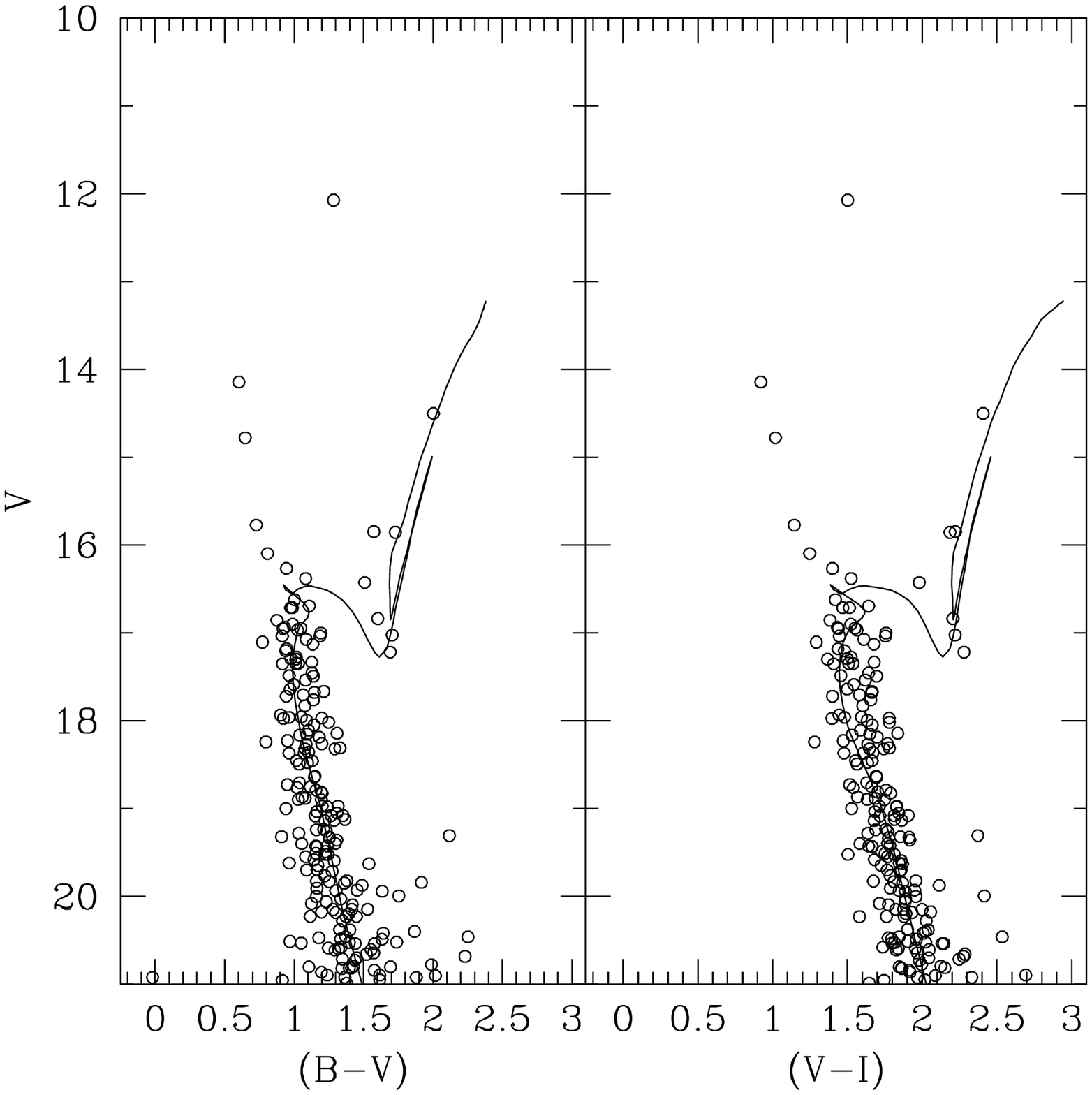,width=\columnwidth}} 
\caption{Isochrone solution for Czernik~32. The isochrone
is for the age of 1 Gyr and metallicity Z=0.008.
The apparent distance modulus is (m-M)=15.7, and the reddening
E(B-V)=0.85 and E(V-I)=1.08. See text for more details. 
Only stars within the derived radius are shown.}
\end{figure} 

\begin{figure} 
\centerline{\psfig{file=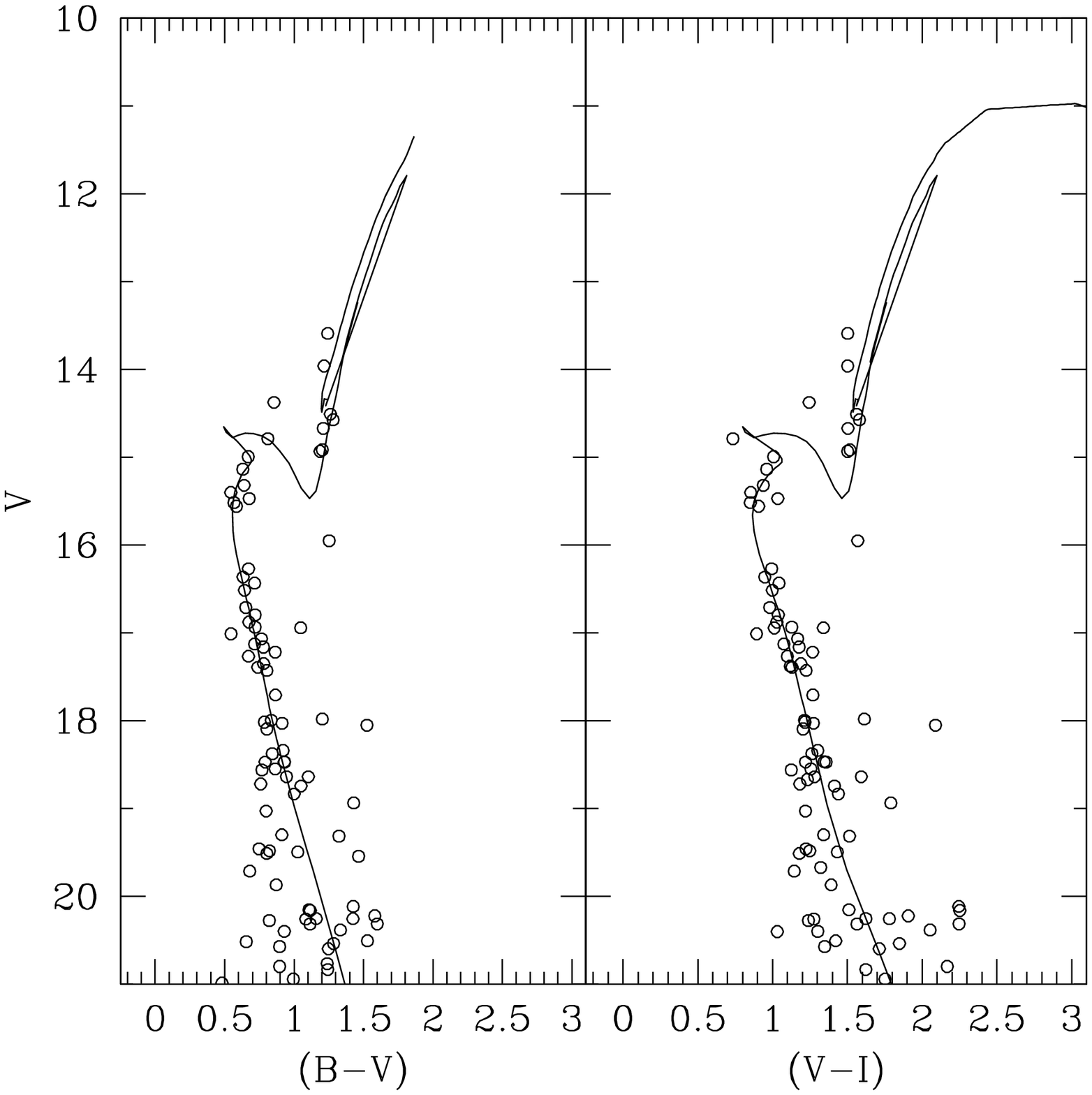,width=\columnwidth}} 
\caption{Isochrone solution for NGC~2225. The isochrone
is for the age of 1.3 Gyrs and metallicity Z=0.008.
The apparent distance modulus is (m-M)=13.5, and the reddening
E(B-V)=0.35 and E(V-I)=0.56. See text for more details. 
Only stars within the derived radius are shown.}
\end{figure}

\begin{figure} 
\centerline{\psfig{file=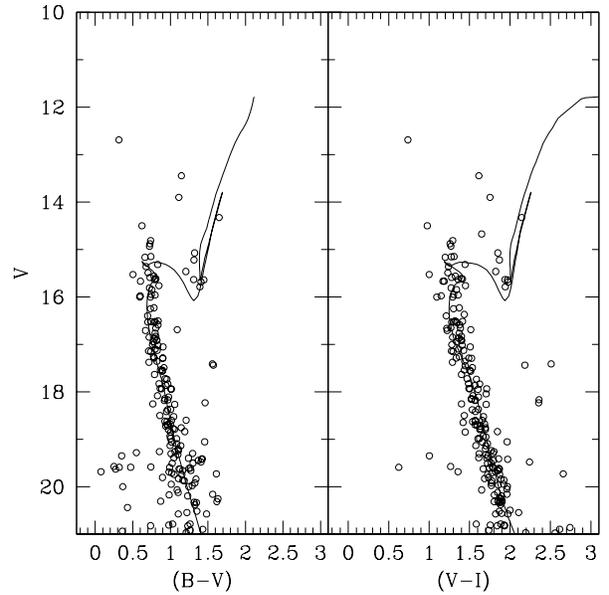,width=\columnwidth}} 
\caption{Isochrone solution for Czernik~32. The isochrone
is for the age of 1 Gyr and metallicity Z=0.008.
The apparent distance modulus is (m-M)=14.5, and the reddening
E(B-V)=0.55 and E(V-I)=0.72. See text for more details. 
Only stars within the derived radius are shown.}
\end{figure}

\section{Conclusions}
We have presented the first CCD $BVI$ photometric study of the 
star clusters Ruprecht~61, Czernik~32, NGC~2225 and NGC~2262. 
Through a star count analysis we have refined previous estimates
of the clusters' coordinates and apparent radii.
A detailed comparison of the clusters' CMDs with
theoretical isochrones has allowed us to infer
the aggregates' basic parameters and their uncertainties
, which
are summarized in Table~4.\\
\noindent
In detail, the fundamental findings of this paper are:

\begin{description}
\item $\bullet$ the best fit reddening estimates support
within the errors a normal extinction law toward the four clusters;\\
\item $\bullet$ all the clusters turn out to be of intermediate age,
and not far from the Sun toward the anti-center direction.\\
\item $\bullet$ the photometric estimates of the metallicity
are lower than solar, as expected
for clusters located between 9 and 11 kpc from the Galactic Center
(Carraro et al. 1998).
\end{description}
 
\begin{table*}
\caption{Fundamental parameters of the studied clusters. The coordinates system is centered on the Sun, with the X and Y axes lying on the Galactic plane
ansd Z perpendicular to the plane.
Y points in the direction of the Galactic rotation, being positive in the first and second Galactic quadrants; Y points toward the Galactic ancticenter;
and Z indicates the north Galactic pole (Lynga 1982).}
\fontsize{8} {9pt}\selectfont
\begin{tabular}{cccccccccccc}
\hline
\multicolumn{1}{c} {$Name$} &
\multicolumn{1}{c} {$Radius(^{\prime})$} &
\multicolumn{1}{c} {$E(B-V)$}  &
\multicolumn{1}{c} {$E(V-I)$}  &
\multicolumn{1}{c} {$(m-M)$} &
\multicolumn{1}{c} {$d_{\odot}$} &
\multicolumn{1}{c} {$X(kpc)$} &
\multicolumn{1}{c} {$Y(kpc)$} &
\multicolumn{1}{c} {$Z(kpc)$} &
\multicolumn{1}{c} {$R_{GC}(kpc)$} &
\multicolumn{1}{c} {$Age(Gyr)$} & 
\multicolumn{1}{c} {Metallicity}\\
\hline
Ruprecht~61 &  1.0 & 0.30$\pm$0.1 & 0.41$\pm$0.1  & 13.85$\pm$0.2 & 3.9 &-3.8 & 1.1 &  0.19 &  9.4 & 1.3$\pm$0.2 & 0.008$\pm$0.002\\
Czernik~32  &  1.0 & 0.85$\pm$0.1 & 1.08$\pm$0.1  & 15.70$\pm$0.2 & 4.1 &-1.7 & 3.7 & -0.12 & 10.8 & 1.0$\pm$0.3 & 0.008$\pm$0.002\\
NGC~2225    &  1.2 & 0.35$\pm$0.1 & 0.50$\pm$0.1  & 13.60$\pm$0.2 & 3.2 &-1.9 & 2.5 & -0.55 & 11.2 & 1.3$\pm$0.2 & 0.008$\pm$0.003\\
NGC~2262    &  1.0 & 0.55$\pm$0.1 & 0.72$\pm$0.1  & 14.50$\pm$0.2 & 3.6 &-1.8 & 3.1 & -0.13 & 11.7 & 1.0$\pm$0.2 & 0.008$\pm$0.004\\
\hline
\end{tabular}
\end{table*}

\section*{Acknowledgements} 
The observations presented in this paper have been carried out at 
Cerro Tololo Interamerican Observatory CTIO (Chile).
CTIO is operated by the Association of Universities for Research in Astronomy,
Inc. (AURA), under a cooperative agreement with the National Science Foundation
as part of the National Optical Astronomy Observatory (NOAO).
The work of G.C. is supported by {\it Fundaci\'on Andes}.
D.G. gratefully acknowledges support from the Chilean
{\sl Centro de Astrof\'\i sica} FONDAP No. 15010003.
This work has been also developed in the framework of 
the {\it Programa Cient\'ifico-Tecnol\'ogico Argentino-Italiano SECYT-MAE
C\'odigo: IT/PA03 - UIII/077 - per\'iodo 2004-2005}.
A.M. acknowledges support from FCT (Portugal) through grant
SFRH/BPD/19105/2004.
This study made use of Simbad and WEBDA databases.

\end{document}